\documentclass{article}[10pt,letterpaper]
\usepackage{fancyhdr}
\usepackage{supertabular}
\usepackage{color}
\usepackage{amsmath}
\usepackage{pbox}
\usepackage{amssymb}
\usepackage{amsbsy}
\usepackage{scalefnt}
\usepackage{ifpdf}
\usepackage{graphicx}
\usepackage{multicol}
\usepackage[colorlinks=true, linkcolor=blue, linktocpage=true,urlcolor=blue]{hyperref}
\usepackage{longtable}
\usepackage{needspace}
\usepackage{eso-pic}
\usepackage{mathptmx}
\usepackage{multirow}
\usepackage{lscape}
\usepackage{textcomp}
\usepackage{changepage}
\usepackage{xfrac}
\usepackage{txfonts}
\usepackage{pdfcomment}
\usepackage[absolute]{textpos}
\ifpdf
\fi
\makeatletter
\renewcommand{\l@section}{\@dottedtocline{2}{1em}{0em}}
\renewcommand{\l@subsection}{\@dottedtocline{3}{2.5cm}{0em}}
\renewcommand{\hrulefill}{\leavevmode \leaders \hrule \@height 1pt \hfill \kern\z@}
\makeatother
\renewcommand{\underline}[1]{\begin{tabular}{@{\extracolsep{\fill}}c@{\extracolsep{\fill}}}#1\\[-0.2cm]\hrulefill\end{tabular}}

\newcounter{chappage}
\AddToShipoutPicture{\stepcounter{chappage}}
\fancypagestyle{plain}{\lhead{}\rhead{}}
\fancypagestyle{single}{\lhead{\leftmark \ \\}\rhead{\leftmark \ \\}}
\fancypagestyle{bob}{\lhead{\leftmark -\thechappage \ \\}\rhead{\leftmark -\thechappage \ \\}}
\begin{document}
\fontsize{9}{10}\selectfont
\fontdimen2\font=1.3\fontdimen2\font
\thispagestyle{empty}
\setcounter{page}{1}
\begin{center}

\vspace{0.5cm}
{ \huge Energy Levels of \ensuremath{^{\textnormal{16}}}Be*}\\
\vspace{1.0cm}
{ \normalsize K. Setoodehnia\ensuremath{^{\textnormal{1,2}}} and J. H. Kelley\ensuremath{^{\textnormal{1,3}}}}\\
\vspace{0.2in}
{ \small \it \ensuremath{^{\textnormal{1}}}Triangle Universities Nuclear Laboratory, Duke University,\\
  Durham, North Carolina 27708, USA.\\
  \ensuremath{^{\textnormal{2}}}Department of Physics, Duke University, Durham, North Carolina\\
  27708, USA.\\
  \ensuremath{^{\textnormal{3}}}Department of Physics, North Carolina State University,\\
  Raleigh, North Carolina 27607, USA}\\
\vspace{0.2in}
\end{center}

\setlength{\parindent}{-0.5cm}
\addtolength{\leftskip}{2cm}
\addtolength{\rightskip}{2cm}
{\bf Abstract: }
In this document, experimental nuclear structure data are evaluated for \ensuremath{^{\textnormal{16}}}Be. The details of each reaction populating \ensuremath{^{\textnormal{16}}}Be levels are compiled and evaluated. The combined results provide a set of adopted values that include level energies, spins and parities, level widths, decay types and other nuclear properties.\\

{\bf Cutoff Date: }
Literature available up to April 10, 2024 has been considered; the primary bibliographic source, the NSR database (\href{https://www.nndc.bnl.gov/nsr/nsrlink.jsp?2011Pr03,B}{2011Pr03}) available at Brookhaven National Laboratory web page: www.nndc.bnl.gov/nsr/.\\

{\bf General Policies and Organization of Material: }
See the April 2025 issue of the {\it Nuclear Data Sheets} or \\https://www.nndc.bnl.gov/nds/docs/NDSPolicies.pdf. \\

{\bf Acknowledgements: }
The authors expresses her gratitude to personnel at the National Nuclear Data Center (NNDC) at Brookhaven National Laboratory for facilitating this work.\\

\vfill

* This work is supported by the Office of Nuclear Physics, Office of Science, U.S. Department of Energy under contracts: DE-FG02-97ER41042 {\textminus} North Carolina State University and DE-FG02-97ER41033 {\textminus} Duke University\\

\setlength{\parindent}{+0.5cm}
\addtolength{\leftskip}{-2cm}
\addtolength{\rightskip}{-2cm}
\newpage
\pagestyle{plain}
\setlength{\columnseprule}{1pt}
\setlength{\columnsep}{1cm}
\begin{center}
\underline{\normalsize Index for A=16}
\end{center}
\hspace{.3cm}\raggedright\underline{Nuclide}\hspace{1cm}\underline{Data Type\mbox{\hspace{2.3cm}}}\hspace{2cm}\underline{Page}\hspace{1cm}
\raggedright\underline{Nuclide}\hspace{1cm}\underline{Data Type\mbox{\hspace{2.3cm}}}\hspace{2cm}\underline{Page}
\begin{adjustwidth}{}{0.05\textwidth}
\begin{multicols}{2}
\setcounter{tocdepth}{3}
\renewcommand{\contentsname}{\protect\vspace{-0.8cm}}
\tableofcontents
\end{multicols}
\end{adjustwidth}
\clearpage
\thispagestyle{empty}
\mbox{}
\clearpage
\clearpage
\pagestyle{bob}
\begin{center}
\section[\ensuremath{^{16}_{\ 4}}Be\ensuremath{_{12}^{~}}]{ }
\vspace{-30pt}
\setcounter{chappage}{1}
\subsection[\hspace{-0.2cm}Adopted Levels]{ }
\vspace{-20pt}
\vspace{0.3cm}
\hypertarget{BE0}{{\bf \small \underline{Adopted \hyperlink{16BE_LEVEL}{Levels}}}}\\
\vspace{4pt}
\vspace{8pt}
\parbox[b][0.3cm]{17.7cm}{\addtolength{\parindent}{-0.2in}Q(\ensuremath{\beta^-})=19.82\ensuremath{\times10^{3}} {\it 13}; S(n)=0.96\ensuremath{\times10^{3}} {\it 10}\hspace{0.2in}\href{https://www.nndc.bnl.gov/nsr/nsrlink.jsp?2024Mo02,B}{2024Mo02}}\\
\parbox[b][0.3cm]{17.7cm}{\addtolength{\parindent}{-0.2in}S(2n)={\textminus}840 keV \textit{30} from (\href{https://www.nndc.bnl.gov/nsr/nsrlink.jsp?2024Mo02,B}{2024Mo02}).}\\
\parbox[b][0.3cm]{17.7cm}{\addtolength{\parindent}{-0.2in}Q(\ensuremath{\beta}\ensuremath{^{-}}): From the mass excesses of \ensuremath{^{\textnormal{16}}}Be and \ensuremath{^{\textnormal{16}}}B given in (\href{https://www.nndc.bnl.gov/nsr/nsrlink.jsp?2024Mo02,B}{2024Mo02}) and (\href{https://www.nndc.bnl.gov/nsr/nsrlink.jsp?2021Wa16,B}{2021Wa16}), respectively.}\\
\parbox[b][0.3cm]{17.7cm}{\addtolength{\parindent}{-0.2in}S(n): From S\ensuremath{_{\textnormal{n}}}(\ensuremath{^{\textnormal{16}}}Be)=S\ensuremath{_{\textnormal{2n}}}(\ensuremath{^{\textnormal{16}}}Be){\textminus}S\ensuremath{_{\textnormal{n}}}(\ensuremath{^{\textnormal{15}}}Be), where S\ensuremath{_{\textnormal{n}}}(\ensuremath{^{\textnormal{15}}}Be)={\textminus}1800 keV \textit{100} (\href{https://www.nndc.bnl.gov/nsr/nsrlink.jsp?2013Sn02,B}{2013Sn02}) as recommended by (\href{https://www.nndc.bnl.gov/nsr/nsrlink.jsp?2021Wa16,B}{2021Wa16}).}\\

\vspace{0.385cm}
\parbox[b][0.3cm]{17.7cm}{\addtolength{\parindent}{-0.2in}See general theoretical analysis of \ensuremath{^{\textnormal{16}}}Be in: \href{https://www.nndc.bnl.gov/nsr/nsrlink.jsp?1981Se06,B}{1981Se06}, \href{https://www.nndc.bnl.gov/nsr/nsrlink.jsp?1985Po10,B}{1985Po10}, \href{https://www.nndc.bnl.gov/nsr/nsrlink.jsp?1987Sa15,B}{1987Sa15}, \href{https://www.nndc.bnl.gov/nsr/nsrlink.jsp?2002Ne24,B}{2002Ne24}, \href{https://www.nndc.bnl.gov/nsr/nsrlink.jsp?2006Ko02,B}{2006Ko02}, \href{https://www.nndc.bnl.gov/nsr/nsrlink.jsp?2008Um02,B}{2008Um02}, \href{https://www.nndc.bnl.gov/nsr/nsrlink.jsp?2009Yu07,B}{2009Yu07},}\\
\parbox[b][0.3cm]{17.7cm}{\href{https://www.nndc.bnl.gov/nsr/nsrlink.jsp?2012It04,B}{2012It04}, \href{https://www.nndc.bnl.gov/nsr/nsrlink.jsp?2015Ka02,B}{2015Ka02}, \href{https://www.nndc.bnl.gov/nsr/nsrlink.jsp?2017Lo03,B}{2017Lo03} (see also \href{https://www.nndc.bnl.gov/nsr/nsrlink.jsp?2016LoZU,B}{2016LoZU}), \href{https://www.nndc.bnl.gov/nsr/nsrlink.jsp?2018Fo07,B}{2018Fo07}, \href{https://www.nndc.bnl.gov/nsr/nsrlink.jsp?2018Ca09,B}{2018Ca09}, \href{https://www.nndc.bnl.gov/nsr/nsrlink.jsp?2019Fo09,B}{2019Fo09}, \href{https://www.nndc.bnl.gov/nsr/nsrlink.jsp?2019Ca03,B}{2019Ca03}, \href{https://www.nndc.bnl.gov/nsr/nsrlink.jsp?2020It02,B}{2020It02}, \href{https://www.nndc.bnl.gov/nsr/nsrlink.jsp?2022Yu02,B}{2022Yu02}, \href{https://www.nndc.bnl.gov/nsr/nsrlink.jsp?2022Gu11,B}{2022Gu11},}\\
\parbox[b][0.3cm]{17.7cm}{and \href{https://www.nndc.bnl.gov/nsr/nsrlink.jsp?2023Mu11,B}{2023Mu11}.}\\
\vspace{12pt}
\hypertarget{16BE_LEVEL}{\underline{$^{16}$Be Levels}}\\
\begin{longtable}[c]{ll}
\multicolumn{2}{c}{\underline{Cross Reference (XREF) Flags}}\\
 \\
\hyperlink{BE1}{\texttt{A }}& \ensuremath{^{\textnormal{1}}}H(\ensuremath{^{\textnormal{17}}}B,2p)\\
\hyperlink{BE2}{\texttt{B }}& \ensuremath{^{\textnormal{9}}}Be(\ensuremath{^{\textnormal{17}}}B,\ensuremath{^{\textnormal{16}}}Be)\\
\hyperlink{BE3}{\texttt{C }}& \ensuremath{^{\textnormal{9}}}Be(\ensuremath{^{\textnormal{40}}}Ar,\ensuremath{^{\textnormal{16}}}Be)\\
\end{longtable}
\vspace{-0.5cm}
\begin{longtable}{ccccccc@{\extracolsep{\fill}}c}
\multicolumn{2}{c}{E(level)$^{}$}&J$^{\pi}$$^{{\hyperlink{BE0LEVEL0}{a}}}$&\multicolumn{2}{c}{\ensuremath{\Gamma}$^{}$}&XREF&Comments&\\[-.2cm]
\multicolumn{2}{c}{\hrulefill}&\hrulefill&\multicolumn{2}{c}{\hrulefill}&\hrulefill&\hrulefill&
\endfirsthead
\multicolumn{1}{r@{}}{0}&\multicolumn{1}{@{}l}{}&\multicolumn{1}{l}{0\ensuremath{^{+}}}&\multicolumn{1}{r@{}}{0}&\multicolumn{1}{@{.}l}{32 MeV {\it 8}}&\multicolumn{1}{l}{\texttt{\hyperlink{BE1}{A}\hyperlink{BE2}{b}\ } }&\parbox[t][0.3cm]{11.004661cm}{\raggedright \%2n=100 (\href{https://www.nndc.bnl.gov/nsr/nsrlink.jsp?2024Mo02,B}{2024Mo02},\href{https://www.nndc.bnl.gov/nsr/nsrlink.jsp?2012Sp01,B}{2012Sp01})\vspace{0.1cm}}&\\
&&&&&&\parbox[t][0.3cm]{11.004661cm}{\raggedright XREF: b(0)\vspace{0.1cm}}&\\
&&&&&&\parbox[t][0.3cm]{11.004661cm}{\raggedright E(level),\ensuremath{\Gamma},J\ensuremath{^{\pi}}: From (\href{https://www.nndc.bnl.gov/nsr/nsrlink.jsp?2024Mo02,B}{2024Mo02}).\vspace{0.1cm}}&\\
&&&&&&\parbox[t][0.3cm]{11.004661cm}{\raggedright E(level): (\href{https://www.nndc.bnl.gov/nsr/nsrlink.jsp?2012Sp01,B}{2012Sp01}) reported that the ground state is unbound to 2n decay by\vspace{0.1cm}}&\\
&&&&&&\parbox[t][0.3cm]{11.004661cm}{\raggedright {\ }{\ }{\ }1.35 MeV \textit{10} and has a width of \ensuremath{\Gamma}=0.8 MeV \textit{+1{\textminus}2} (\href{https://www.nndc.bnl.gov/nsr/nsrlink.jsp?2012Sp01,B}{2012Sp01}). However, the\vspace{0.1cm}}&\\
&&&&&&\parbox[t][0.3cm]{11.004661cm}{\raggedright {\ }{\ }{\ }limited energy resolution of this experiment resulted in the observation of a\vspace{0.1cm}}&\\
&&&&&&\parbox[t][0.3cm]{11.004661cm}{\raggedright {\ }{\ }{\ }broad structure comprising the unresolved ground+first excited states. The\vspace{0.1cm}}&\\
&&&&&&\parbox[t][0.3cm]{11.004661cm}{\raggedright {\ }{\ }{\ }superior energy resolution of (\href{https://www.nndc.bnl.gov/nsr/nsrlink.jsp?2024Mo02,B}{2024Mo02}) resolved these states, which are\vspace{0.1cm}}&\\
&&&&&&\parbox[t][0.3cm]{11.004661cm}{\raggedright {\ }{\ }{\ }measured with much higher statistics. (\href{https://www.nndc.bnl.gov/nsr/nsrlink.jsp?2024Mo02,B}{2024Mo02}) reports that the ground state\vspace{0.1cm}}&\\
&&&&&&\parbox[t][0.3cm]{11.004661cm}{\raggedright {\ }{\ }{\ }is unbound to 2n decay by 0.84 MeV \textit{3}.\vspace{0.1cm}}&\\
&&&&&&\parbox[t][0.3cm]{11.004661cm}{\raggedright Decay is through dineutron emission (\href{https://www.nndc.bnl.gov/nsr/nsrlink.jsp?2024Mo02,B}{2024Mo02}, \href{https://www.nndc.bnl.gov/nsr/nsrlink.jsp?2012Sp01,B}{2012Sp01}). A realistic 3-body\vspace{0.1cm}}&\\
&&&&&&\parbox[t][0.3cm]{11.004661cm}{\raggedright {\ }{\ }{\ }modeling by (\href{https://www.nndc.bnl.gov/nsr/nsrlink.jsp?2024Mo02,B}{2024Mo02}) showed that the ground state manifests itself with a\vspace{0.1cm}}&\\
&&&&&&\parbox[t][0.3cm]{11.004661cm}{\raggedright {\ }{\ }{\ }strong dineutron-like configuration.\vspace{0.1cm}}&\\
&&&&&&\parbox[t][0.3cm]{11.004661cm}{\raggedright Mass excess(\ensuremath{^{\textnormal{16}}}Be\ensuremath{_{\textnormal{g.s.}}})=56.93 MeV \textit{13} (\href{https://www.nndc.bnl.gov/nsr/nsrlink.jsp?2024Mo02,B}{2024Mo02}).\vspace{0.1cm}}&\\
&&&&&&\parbox[t][0.3cm]{11.004661cm}{\raggedright Using the deduced mass excess from (\href{https://www.nndc.bnl.gov/nsr/nsrlink.jsp?2024Mo02,B}{2024Mo02}), the evaluator determined the\vspace{0.1cm}}&\\
&&&&&&\parbox[t][0.3cm]{11.004661cm}{\raggedright {\ }{\ }{\ }mass of \ensuremath{^{\textnormal{16}}}Be as 16.06112 u \textit{14}.\vspace{0.1cm}}&\\
\multicolumn{1}{r@{}}{1.31\ensuremath{\times10^{3}}}&\multicolumn{1}{@{ }l}{{\it 6}}&\multicolumn{1}{l}{2\ensuremath{^{+}}}&\multicolumn{1}{r@{}}{0}&\multicolumn{1}{@{.}l}{95 MeV {\it 15}}&\multicolumn{1}{l}{\texttt{\hyperlink{BE1}{A}\hyperlink{BE2}{b}\ } }&\parbox[t][0.3cm]{11.004661cm}{\raggedright \%2n=100 (\href{https://www.nndc.bnl.gov/nsr/nsrlink.jsp?2024Mo02,B}{2024Mo02})\vspace{0.1cm}}&\\
&&&&&&\parbox[t][0.3cm]{11.004661cm}{\raggedright XREF: b(0)\vspace{0.1cm}}&\\
&&&&&&\parbox[t][0.3cm]{11.004661cm}{\raggedright E(level),\ensuremath{\Gamma},J\ensuremath{^{\pi}}: From (\href{https://www.nndc.bnl.gov/nsr/nsrlink.jsp?2024Mo02,B}{2024Mo02}).\vspace{0.1cm}}&\\
&&&&&&\parbox[t][0.3cm]{11.004661cm}{\raggedright E(level): The first excited state is unbound to 2n decay by 2.15 MeV \textit{5}.\vspace{0.1cm}}&\\
&&&&&&\parbox[t][0.3cm]{11.004661cm}{\raggedright (\href{https://www.nndc.bnl.gov/nsr/nsrlink.jsp?2024Mo02,B}{2024Mo02}): This state decays by dineutron emission. As a result of a realistic\vspace{0.1cm}}&\\
&&&&&&\parbox[t][0.3cm]{11.004661cm}{\raggedright {\ }{\ }{\ }3-body modeling, it was found that this state exhibits a diffuse n-n spatial\vspace{0.1cm}}&\\
&&&&&&\parbox[t][0.3cm]{11.004661cm}{\raggedright {\ }{\ }{\ }distribution.\vspace{0.1cm}}&\\
\end{longtable}
\parbox[b][0.3cm]{17.7cm}{\makebox[1ex]{\ensuremath{^{\hypertarget{BE0LEVEL0}{a}}}} From shell model calculations by (\href{https://www.nndc.bnl.gov/nsr/nsrlink.jsp?2024Mo02,B}{2024Mo02}).}\\
\vspace{0.5cm}
\clearpage
\subsection[\hspace{-0.2cm}\ensuremath{^{\textnormal{1}}}H(\ensuremath{^{\textnormal{17}}}B,2p)]{ }
\vspace{-27pt}
\vspace{0.3cm}
\hypertarget{BE1}{{\bf \small \underline{\ensuremath{^{\textnormal{1}}}H(\ensuremath{^{\textnormal{17}}}B,2p)\hspace{0.2in}\href{https://www.nndc.bnl.gov/nsr/nsrlink.jsp?2024Mo02,B}{2024Mo02}}}}\\
\vspace{4pt}
\vspace{8pt}
\parbox[b][0.3cm]{17.7cm}{\addtolength{\parindent}{-0.2in}\href{https://www.nndc.bnl.gov/nsr/nsrlink.jsp?2024Mo02,B}{2024Mo02}: The authors populated the ground and first excited states of \ensuremath{^{\textnormal{16}}}Be by proton knockout from a \ensuremath{^{\textnormal{17}}}B beam and}\\
\parbox[b][0.3cm]{17.7cm}{investigated their structure and decay by measuring the \ensuremath{^{\textnormal{14}}}Be+2n decay products using an experimental configuration which offered}\\
\parbox[b][0.3cm]{17.7cm}{better energy resolution and an improved acceptance relative to the measurement of (\href{https://www.nndc.bnl.gov/nsr/nsrlink.jsp?2012Sp01,B}{2012Sp01}).}\\
\parbox[b][0.3cm]{17.7cm}{\addtolength{\parindent}{-0.2in}A \ensuremath{^{\textnormal{17}}}B beam with E\ensuremath{\sim}277 MeV/nucleon was produced from fragmentation of a \ensuremath{^{\textnormal{48}}}Ca beam on a thick Be target at the RIBF facility}\\
\parbox[b][0.3cm]{17.7cm}{in RIKEN. The \ensuremath{^{\textnormal{17}}}B beam was purified using the BigRIPS fragment separator and impinged on a 15-cm-thick liquid hydrogen}\\
\parbox[b][0.3cm]{17.7cm}{target at the MINOS target position. The \ensuremath{^{\textnormal{16}}}Be nuclei were produced via the \ensuremath{^{\textnormal{1}}}H(\ensuremath{^{\textnormal{17}}}Be,2p) reaction. The protons were measured}\\
\parbox[b][0.3cm]{17.7cm}{using the MINOS TPC. The \ensuremath{^{\textnormal{16}}}Be nuclei decayed in flight, and the \ensuremath{^{\textnormal{14}}}Be and neutrons from this decay were momentum analyzed}\\
\parbox[b][0.3cm]{17.7cm}{using the SAMURAI spectrometer and the NEBULA array, respectively.}\\
\parbox[b][0.3cm]{17.7cm}{\addtolength{\parindent}{-0.2in}The reaction vertex was reconstructed with a 5-mm position resolution using the two reaction protons$'$ trajectories and}\\
\parbox[b][0.3cm]{17.7cm}{event-by-event beam tracking. The excitation spectrum was obtained from the reconstruction of the relative energy of \ensuremath{^{\textnormal{14}}}Be+n+n}\\
\parbox[b][0.3cm]{17.7cm}{decay products using an invariant mass analysis. The ground and first excited states were observed with high statistics and are well}\\
\parbox[b][0.3cm]{17.7cm}{resolved. The mass excess of \ensuremath{^{\textnormal{16}}}Be\ensuremath{_{\textnormal{g.s.}}} was deduced. Some excess counts were measured above \ensuremath{\sim}4 MeV, which were attributed to a}\\
\parbox[b][0.3cm]{17.7cm}{non-resonant continuum or broad, weakly populated higher-lying structures.}\\
\parbox[b][0.3cm]{17.7cm}{\addtolength{\parindent}{-0.2in}A shell model calculation was performed using the WBP Hamiltonian, and the predictions are in good agreement with the}\\
\parbox[b][0.3cm]{17.7cm}{experimental results. The decays of the observed \ensuremath{^{\textnormal{16}}}Be states were investigated using Dalitz plots supported by simulations. No}\\
\parbox[b][0.3cm]{17.7cm}{evidence for sequential decay via \ensuremath{^{\textnormal{15}}}Be states was observed. Both the populated states decay via direct 2n emission. A 3-body}\\
\parbox[b][0.3cm]{17.7cm}{modeling of \ensuremath{^{\textnormal{14}}}Be+n+n was performed to study the nature of these decays. The authors emphasized the importance of using realistic}\\
\parbox[b][0.3cm]{17.7cm}{wave functions that evolve with time to describe the decay. Such a treatment was missing in (\href{https://www.nndc.bnl.gov/nsr/nsrlink.jsp?2012Sp01,B}{2012Sp01}). Moreover, the earlier}\\
\parbox[b][0.3cm]{17.7cm}{work did not consider the n-n final state interaction, which oversimplified their 2-body (\ensuremath{^{\textnormal{14}}}Be+2n) decay study. The results of the}\\
\parbox[b][0.3cm]{17.7cm}{present 3-body decay are in good agreement (except the predicted widths) with the experimental ones and support the dineutron}\\
\parbox[b][0.3cm]{17.7cm}{emission from both of the observed states.}\\
\vspace{12pt}
\underline{$^{16}$Be Levels}\\
\begin{longtable}{cccccc@{\extracolsep{\fill}}c}
\multicolumn{2}{c}{E(level)$^{}$}&J$^{\pi}$$^{{\hyperlink{BE1LEVEL0}{a}}}$&\multicolumn{2}{c}{\ensuremath{\Gamma}$^{}$}&Comments&\\[-.2cm]
\multicolumn{2}{c}{\hrulefill}&\hrulefill&\multicolumn{2}{c}{\hrulefill}&\hrulefill&
\endfirsthead
\multicolumn{1}{r@{}}{0}&\multicolumn{1}{@{}l}{}&\multicolumn{1}{l}{0\ensuremath{^{+}}}&\multicolumn{1}{r@{}}{0}&\multicolumn{1}{@{.}l}{32 MeV {\it 8}}&\parbox[t][0.3cm]{12.22428cm}{\raggedright \%2n=100\vspace{0.1cm}}&\\
&&&&&\parbox[t][0.3cm]{12.22428cm}{\raggedright E(level): The ground state is unbound to 2n decay by 0.84 MeV \textit{3}. Decay is through\vspace{0.1cm}}&\\
&&&&&\parbox[t][0.3cm]{12.22428cm}{\raggedright {\ }{\ }{\ }dineutron emission. A realistic 3-body modeling showed that the ground state manifests\vspace{0.1cm}}&\\
&&&&&\parbox[t][0.3cm]{12.22428cm}{\raggedright {\ }{\ }{\ }itself with a strong dineutron-like configuration.\vspace{0.1cm}}&\\
&&&&&\parbox[t][0.3cm]{12.22428cm}{\raggedright Using S\ensuremath{_{\textnormal{2n}}}(\ensuremath{^{\textnormal{16}}}Be) and mass excess(\ensuremath{^{\textnormal{14}}}Be)=39.95 MeV \textit{13} from (\href{https://www.nndc.bnl.gov/nsr/nsrlink.jsp?2021Wa16,B}{2021Wa16}), the mass\vspace{0.1cm}}&\\
&&&&&\parbox[t][0.3cm]{12.22428cm}{\raggedright {\ }{\ }{\ }excess of \ensuremath{^{\textnormal{16}}}Be\ensuremath{_{\textnormal{g.s.}}} was deduced to be 56.93 MeV \textit{13}. The uncertainty is dominated by\vspace{0.1cm}}&\\
&&&&&\parbox[t][0.3cm]{12.22428cm}{\raggedright {\ }{\ }{\ }the mass excess of \ensuremath{^{\textnormal{14}}}Be.\vspace{0.1cm}}&\\
&&&&&\parbox[t][0.3cm]{12.22428cm}{\raggedright Using the deduced mass excess from (\href{https://www.nndc.bnl.gov/nsr/nsrlink.jsp?2024Mo02,B}{2024Mo02}), the evaluator determined the mass of\vspace{0.1cm}}&\\
&&&&&\parbox[t][0.3cm]{12.22428cm}{\raggedright {\ }{\ }{\ }\ensuremath{^{\textnormal{16}}}Be as 16.06112 u \textit{14}. These results supersede the earlier results of (\href{https://www.nndc.bnl.gov/nsr/nsrlink.jsp?2012Sp01,B}{2012Sp01}), where\vspace{0.1cm}}&\\
&&&&&\parbox[t][0.3cm]{12.22428cm}{\raggedright {\ }{\ }{\ }the ground and first excited states were apparently unresolved.\vspace{0.1cm}}&\\
\multicolumn{1}{r@{}}{1.31\ensuremath{\times10^{3}}}&\multicolumn{1}{@{ }l}{{\it 6}}&\multicolumn{1}{l}{2\ensuremath{^{+}}}&\multicolumn{1}{r@{}}{0}&\multicolumn{1}{@{.}l}{95 MeV {\it 15}}&\parbox[t][0.3cm]{12.22428cm}{\raggedright \%2n=100\vspace{0.1cm}}&\\
&&&&&\parbox[t][0.3cm]{12.22428cm}{\raggedright E(level): The first state is unbound to 2n decay by 2.15 MeV \textit{5}. This state decays by\vspace{0.1cm}}&\\
&&&&&\parbox[t][0.3cm]{12.22428cm}{\raggedright {\ }{\ }{\ }dineutron emission. As a result of a realistic 3-body modeling, this state exhibits a\vspace{0.1cm}}&\\
&&&&&\parbox[t][0.3cm]{12.22428cm}{\raggedright {\ }{\ }{\ }diffuse n-n spatial distribution.\vspace{0.1cm}}&\\
\end{longtable}
\parbox[b][0.3cm]{17.7cm}{\makebox[1ex]{\ensuremath{^{\hypertarget{BE1LEVEL0}{a}}}} From shell model calculations by (\href{https://www.nndc.bnl.gov/nsr/nsrlink.jsp?2024Mo02,B}{2024Mo02}).}\\
\vspace{0.5cm}
\clearpage
\subsection[\hspace{-0.2cm}\ensuremath{^{\textnormal{9}}}Be(\ensuremath{^{\textnormal{17}}}B,\ensuremath{^{\textnormal{16}}}Be)]{ }
\vspace{-27pt}
\vspace{0.3cm}
\hypertarget{BE2}{{\bf \small \underline{\ensuremath{^{\textnormal{9}}}Be(\ensuremath{^{\textnormal{17}}}B,\ensuremath{^{\textnormal{16}}}Be)\hspace{0.2in}\href{https://www.nndc.bnl.gov/nsr/nsrlink.jsp?2012Sp01,B}{2012Sp01}}}}\\
\vspace{4pt}
\vspace{8pt}
\parbox[b][0.3cm]{17.7cm}{\addtolength{\parindent}{-0.2in}\href{https://www.nndc.bnl.gov/nsr/nsrlink.jsp?2012Sp01,B}{2012Sp01}: The authors populated a broad state associated with the ground state of \ensuremath{^{\textnormal{16}}}Be by fragmenting \ensuremath{^{\textnormal{17}}}B nuclei. They studied}\\
\parbox[b][0.3cm]{17.7cm}{\ensuremath{^{\textnormal{16}}}Be decay by measuring complete \ensuremath{^{\textnormal{14}}}Be+2n kinematics. The aim was to determine the \ensuremath{^{\textnormal{16}}}Be mass and evaluate n-n correlations in}\\
\parbox[b][0.3cm]{17.7cm}{search of dineutron decay.}\\
\parbox[b][0.3cm]{17.7cm}{\addtolength{\parindent}{-0.2in}The \ensuremath{^{\textnormal{16}}}Be nuclei were formed in a 2 step process: first a 120 MeV/nucleon \ensuremath{^{\textnormal{22}}}Ne beam was fragmented in a 2938 mg/cm\ensuremath{^{\textnormal{2}}} Be}\\
\parbox[b][0.3cm]{17.7cm}{target to produce \ensuremath{^{\textnormal{17}}}B ions that were purified in the A1900 at the MSU/NSCL, second the \ensuremath{^{\textnormal{17}}}B beam at 53 MeV/nucleon impinged}\\
\parbox[b][0.3cm]{17.7cm}{on a 470 mg/cm\ensuremath{^{\textnormal{2}}} \ensuremath{^{\textnormal{9}}}Be target where the \ensuremath{^{\textnormal{16}}}Be nuclei were formed by fragmentation.}\\
\parbox[b][0.3cm]{17.7cm}{\addtolength{\parindent}{-0.2in}The \ensuremath{^{\textnormal{16}}}Be nuclei decayed in flight and the residual \ensuremath{^{\textnormal{14}}}Be+2n were momentum analyzed using the 43\ensuremath{^\circ} Sweeper dipole magnet and}\\
\parbox[b][0.3cm]{17.7cm}{the MONA array. Kinematic energy reconstruction indicated the particle unbound \ensuremath{^{\textnormal{16}}}Be ground state is at E\ensuremath{_{\textnormal{rel}}}(\ensuremath{^{\textnormal{14}}}Be+2n)=1.35 MeV}\\
\parbox[b][0.3cm]{17.7cm}{\textit{10}. Further analysis of the \ensuremath{^{\textnormal{14}}}Be+n and n+n energy and angular correlations were consistent with dineutron emission from \ensuremath{^{\textnormal{16}}}Be,}\\
\parbox[b][0.3cm]{17.7cm}{and were inconsistent with either sequential decay through \ensuremath{^{\textnormal{15}}}Be or simultaneous 3-body breakup into the \ensuremath{^{\textnormal{14}}}Be+n+n continuum.}\\
\parbox[b][0.3cm]{17.7cm}{See also (\href{https://www.nndc.bnl.gov/nsr/nsrlink.jsp?2013Th04,B}{2013Th04}).}\\
\parbox[b][0.3cm]{17.7cm}{\addtolength{\parindent}{-0.2in}The most recent study by (\href{https://www.nndc.bnl.gov/nsr/nsrlink.jsp?2024Mo02,B}{2024Mo02}) finds a similar spectrum to that of Fig. 2a in (\href{https://www.nndc.bnl.gov/nsr/nsrlink.jsp?2012Sp01,B}{2012Sp01}). As a result of much higher}\\
\parbox[b][0.3cm]{17.7cm}{statistics and a better energy resolution achieved by (\href{https://www.nndc.bnl.gov/nsr/nsrlink.jsp?2024Mo02,B}{2024Mo02}), it is apparent that (\href{https://www.nndc.bnl.gov/nsr/nsrlink.jsp?2012Sp01,B}{2012Sp01}) observed the unresolved}\\
\parbox[b][0.3cm]{17.7cm}{ground+first excited states of \ensuremath{^{\textnormal{16}}}Be.}\\
\vspace{12pt}
\underline{$^{16}$Be Levels}\\
\begin{longtable}{cccccc@{\extracolsep{\fill}}c}
\multicolumn{2}{c}{E(level)$^{{\hyperlink{BE2LEVEL0}{a}}}$}&J$^{\pi}$$^{}$&\multicolumn{2}{c}{\ensuremath{\Gamma}$^{}$}&Comments&\\[-.2cm]
\multicolumn{2}{c}{\hrulefill}&\hrulefill&\multicolumn{2}{c}{\hrulefill}&\hrulefill&
\endfirsthead
\multicolumn{1}{r@{}}{0}&\multicolumn{1}{@{}l}{}&\multicolumn{1}{l}{0\ensuremath{^{+}}}&\multicolumn{1}{r@{}}{0}&\multicolumn{1}{@{.}l}{8 MeV {\it 2}}&\parbox[t][0.3cm]{12.97542cm}{\raggedright \%2n=100\vspace{0.1cm}}&\\
&&&&&\parbox[t][0.3cm]{12.97542cm}{\raggedright E(level): A broad \ensuremath{\Gamma}=0.8 MeV \textit{+1{\textminus}2} group is reported to dineutron decay with\vspace{0.1cm}}&\\
&&&&&\parbox[t][0.3cm]{12.97542cm}{\raggedright {\ }{\ }{\ }E(\ensuremath{^{\textnormal{14}}}Be+2n)=1.35 MeV \textit{10}. Subsequent results by (\href{https://www.nndc.bnl.gov/nsr/nsrlink.jsp?2024Mo02,B}{2024Mo02}) suggest this group is the\vspace{0.1cm}}&\\
&&&&&\parbox[t][0.3cm]{12.97542cm}{\raggedright {\ }{\ }{\ }unresolved ground+first excited states of \ensuremath{^{\textnormal{16}}}Be. Therefore, the results of (\href{https://www.nndc.bnl.gov/nsr/nsrlink.jsp?2012Sp01,B}{2012Sp01}) were not\vspace{0.1cm}}&\\
&&&&&\parbox[t][0.3cm]{12.97542cm}{\raggedright {\ }{\ }{\ }used for the Adopted dataset.\vspace{0.1cm}}&\\
\end{longtable}
\parbox[b][0.3cm]{17.7cm}{\makebox[1ex]{\ensuremath{^{\hypertarget{BE2LEVEL0}{a}}}} Unresolved ground+first excited states.}\\
\vspace{0.5cm}
\clearpage
\subsection[\hspace{-0.2cm}\ensuremath{^{\textnormal{9}}}Be(\ensuremath{^{\textnormal{40}}}Ar,\ensuremath{^{\textnormal{16}}}Be)]{ }
\vspace{-27pt}
\vspace{0.3cm}
\hypertarget{BE3}{{\bf \small \underline{\ensuremath{^{\textnormal{9}}}Be(\ensuremath{^{\textnormal{40}}}Ar,\ensuremath{^{\textnormal{16}}}Be)\hspace{0.2in}\href{https://www.nndc.bnl.gov/nsr/nsrlink.jsp?2003Ba47,B}{2003Ba47}}}}\\
\vspace{4pt}
\vspace{8pt}
\parbox[b][0.3cm]{17.7cm}{\addtolength{\parindent}{-0.2in}\href{https://www.nndc.bnl.gov/nsr/nsrlink.jsp?2003Ba47,B}{2003Ba47}: The authors analyzed the \ensuremath{^{\textnormal{40}}}Ar+\ensuremath{^{\textnormal{9}}}Be fragmentation products in search of evidence for particle bound states in \ensuremath{^{\textnormal{16}}}Be.}\\
\parbox[b][0.3cm]{17.7cm}{\addtolength{\parindent}{-0.2in}A beam of 140 MeV/nucleon \ensuremath{^{\textnormal{40}}}Ar ions, from the NSCL coupled cyclotron facility, impinged on a 1.5 g/cm\ensuremath{^{\textnormal{2}}} \ensuremath{^{\textnormal{nat}}}Be target. The}\\
\parbox[b][0.3cm]{17.7cm}{resulting fragmentation products were momentum analyzed using the A1900 fragment separator. The products were detected using a}\\
\parbox[b][0.3cm]{17.7cm}{position sensitive PPAC, a 500 \ensuremath{\mu}m thick Si \ensuremath{\Delta}E detector and a stopping thickness plastic E scintillator that were located at the}\\
\parbox[b][0.3cm]{17.7cm}{final focal plane of the device. The time difference between a thin plastic scintillator located at the intermediate image of the}\\
\parbox[b][0.3cm]{17.7cm}{separator and the thick stopping detector were compared to determine the time-of-flight (ToF) between the two image planes. The}\\
\parbox[b][0.3cm]{17.7cm}{particle identification at the focal plane was determined using both \ensuremath{\Delta}E-E and \ensuremath{\Delta}E-ToF techniques.}\\
\parbox[b][0.3cm]{17.7cm}{\addtolength{\parindent}{-0.2in}No events corresponding to \ensuremath{^{\textnormal{16}}}Be were observed. By comparison, \ensuremath{^{\textnormal{6,8}}}He, \ensuremath{^{\textnormal{9,11}}}Li, \ensuremath{^{\textnormal{12,14}}}Be, \ensuremath{^{\textnormal{17,19}}}B and \ensuremath{^{\textnormal{20}}}C nuclides were observed}\\
\parbox[b][0.3cm]{17.7cm}{at the focal plane. The measured intensity of \ensuremath{^{\textnormal{19}}}B was expected to be an order of magnitude lower than that of \ensuremath{^{\textnormal{16}}}Be. As a result,}\\
\parbox[b][0.3cm]{17.7cm}{the authors conclude \ensuremath{^{\textnormal{16}}}Be is unstable to neutron emission. See also (\href{https://www.nndc.bnl.gov/nsr/nsrlink.jsp?2004Th15,B}{2004Th15}).}\\
\vspace{12pt}
\end{center}
\clearpage
\newpage
\pagestyle{plain}
\section[References]{ }
\vspace{-30pt}
\begin{longtable}{l@{\hskip 0.9cm}l}
\multicolumn{2}{c}{REFERENCES FOR A=16}\\
&\endfirsthead
\multicolumn{2}{c}{REFERENCES FOR A=16(CONTINUED)}\\
&\endhead
\href{https://www.nndc.bnl.gov/nsr/nsrlink.jsp?1981Se06,B}{1981Se06}&\parbox[t]{6in}{\addtolength{\parindent}{-0.25cm}M.Seya, M.Kohno, S.Nagata - Prog.Theor.Phys.(Kyoto) 65, 204 (1981).}\\
&\parbox[t]{6in}{\addtolength{\parindent}{-0.25cm} \textit{Nuclear Binding Mechanism and Structure of Neutron-Rich Be and B Isotopes by Molecular-Orbital Model.}}\\
\href{https://www.nndc.bnl.gov/nsr/nsrlink.jsp?1985Po10,B}{1985Po10}&\parbox[t]{6in}{\addtolength{\parindent}{-0.25cm}N.A.F.M.Poppelier, L.D.Wood, P.W.M.Glaudemans - Phys.Lett. 157B, 120 (1985).}\\
&\parbox[t]{6in}{\addtolength{\parindent}{-0.25cm} \textit{Properties of Exotic p-Shell Nuclei.}}\\
\href{https://www.nndc.bnl.gov/nsr/nsrlink.jsp?1987Sa15,B}{1987Sa15}&\parbox[t]{6in}{\addtolength{\parindent}{-0.25cm}H.Sagawa, H.Toki - J.Phys.(London) G13, 453 (1987).}\\
&\parbox[t]{6in}{\addtolength{\parindent}{-0.25cm} \textit{Hartree-Fock Calculations of Light Neutron-Rich Nuclei.}}\\
\href{https://www.nndc.bnl.gov/nsr/nsrlink.jsp?2002Ne24,B}{2002Ne24}&\parbox[t]{6in}{\addtolength{\parindent}{-0.25cm}V.A.Nesterov - Iader.Fiz.Enerh. 3 no.1, 22 (2002).}\\
&\parbox[t]{6in}{\addtolength{\parindent}{-0.25cm} \textit{Application of the extended Tomas {\textminus} Fermi method for investigation ofproperties of light atomic nuclei with neutron excess.}}\\
\href{https://www.nndc.bnl.gov/nsr/nsrlink.jsp?2003Ba47,B}{2003Ba47}&\parbox[t]{6in}{\addtolength{\parindent}{-0.25cm}T.Baumann, N.Frank, B.A.Luther, D.J.Morrissey et al. - Phys.Rev. C 67, 061303 (2003).}\\
&\parbox[t]{6in}{\addtolength{\parindent}{-0.25cm} \textit{First search for \ensuremath{^{\textnormal{16}}}Be.}}\\
\href{https://www.nndc.bnl.gov/nsr/nsrlink.jsp?2004Th15,B}{2004Th15}&\parbox[t]{6in}{\addtolength{\parindent}{-0.25cm}M.Thoennessen - Acta Phys.Hung.N.S. 21, 379 (2004).}\\
&\parbox[t]{6in}{\addtolength{\parindent}{-0.25cm} \textit{Pushing the Limits of Nuclear Stability.}}\\
\href{https://www.nndc.bnl.gov/nsr/nsrlink.jsp?2006Ko02,B}{2006Ko02}&\parbox[t]{6in}{\addtolength{\parindent}{-0.25cm}V.B.Kopeliovich, A.M.Shunderuk, G.K.Matushko - Phys.Atomic Nuclei 69, 120 (2006).}\\
&\parbox[t]{6in}{\addtolength{\parindent}{-0.25cm} \textit{Mass Splittings of Nuclear Isotopes in Chiral Soliton Approach.}}\\
\href{https://www.nndc.bnl.gov/nsr/nsrlink.jsp?2008Um02,B}{2008Um02}&\parbox[t]{6in}{\addtolength{\parindent}{-0.25cm}A.Umeya, G.Kaneko, T.Haneda, K.Muto - Phys.Rev. C 77, 044301 (2008).}\\
&\parbox[t]{6in}{\addtolength{\parindent}{-0.25cm} \textit{Enhancement of neutron quadrupole motion beyond N = 8.}}\\
\href{https://www.nndc.bnl.gov/nsr/nsrlink.jsp?2009Yu07,B}{2009Yu07}&\parbox[t]{6in}{\addtolength{\parindent}{-0.25cm}C.-X.Yuan, C.Qi, F.-R.Xu - Chin.Phys.C 33, Supplement 1, 55 (2009).}\\
&\parbox[t]{6in}{\addtolength{\parindent}{-0.25cm} \textit{Shell-model studies of the N = 14 and 16 shell closures in neutron-rich nuclei.}}\\
\href{https://www.nndc.bnl.gov/nsr/nsrlink.jsp?2011Pr03,B}{2011Pr03}&\parbox[t]{6in}{\addtolength{\parindent}{-0.25cm}B.Pritychenko, E.Betak, M.A.Kellett, B.Singh, J.Totans - Nucl.Instrum.Methods Phys.Res. A640, 213 (2011).}\\
&\parbox[t]{6in}{\addtolength{\parindent}{-0.25cm} \textit{The Nuclear Science References (NSR) database and Web Retrieval System.}}\\
\href{https://www.nndc.bnl.gov/nsr/nsrlink.jsp?2012It04,B}{2012It04}&\parbox[t]{6in}{\addtolength{\parindent}{-0.25cm}M.Ito - Prog.Theor.Phys.(Kyoto), Suppl. 196, 289 (2012).}\\
&\parbox[t]{6in}{\addtolength{\parindent}{-0.25cm} \textit{Studies of Light Neutron-Excess Nuclei from Bounds to Continuum.}}\\
\href{https://www.nndc.bnl.gov/nsr/nsrlink.jsp?2012Sp01,B}{2012Sp01}&\parbox[t]{6in}{\addtolength{\parindent}{-0.25cm}A.Spyrou, Z.Kohley, T.Baumann, D.Bazin et al. - Phys.Rev.Lett. 108, 102501 (2012).}\\
&\parbox[t]{6in}{\addtolength{\parindent}{-0.25cm} \textit{First Observation of Ground State Dineutron Decay: \ensuremath{^{\textnormal{16}}}Be.}}\\
\href{https://www.nndc.bnl.gov/nsr/nsrlink.jsp?2013Sn02,B}{2013Sn02}&\parbox[t]{6in}{\addtolength{\parindent}{-0.25cm}J.Snyder, T.Baumann, G.Christian, R.A.Haring-Kaye et al. - Phys.Rev. C 88, 031303 (2013).}\\
&\parbox[t]{6in}{\addtolength{\parindent}{-0.25cm} \textit{First observation of \ensuremath{^{\textnormal{15}}}Be.}}\\
\href{https://www.nndc.bnl.gov/nsr/nsrlink.jsp?2013Th04,B}{2013Th04}&\parbox[t]{6in}{\addtolength{\parindent}{-0.25cm}M.Thoennessen, Z.Kohley, A.Spyrou, E.Lunderberg et al. - Acta Phys.Pol. B44, 543 (2013).}\\
&\parbox[t]{6in}{\addtolength{\parindent}{-0.25cm} \textit{Observation of Ground-state Two-neutron Decay.}}\\
\href{https://www.nndc.bnl.gov/nsr/nsrlink.jsp?2015Ka02,B}{2015Ka02}&\parbox[t]{6in}{\addtolength{\parindent}{-0.25cm}Y.Kanada-Enyo - Phys.Rev. C 91, 014315 (2015).}\\
&\parbox[t]{6in}{\addtolength{\parindent}{-0.25cm} \textit{Proton radii of Be, B, and C isotopes.}}\\
\href{https://www.nndc.bnl.gov/nsr/nsrlink.jsp?2016LoZU,B}{2016LoZU}&\parbox[t]{6in}{\addtolength{\parindent}{-0.25cm}A.E.Lovell, F.M.Nunes, I.J.Thompson - Proc.21sth Int.Conf. on Few Body Problems in Physics, Chicago, IL, USA, May 18-22, 2015, C.Elster,et al. (Eds.), p.06015 (2016);EPJ Web of Conf.Vol. 113 (2016).}\\
&\parbox[t]{6in}{\addtolength{\parindent}{-0.25cm} \textit{Two neutron decay of \ensuremath{^{\textnormal{16}}}Be.}}\\
\href{https://www.nndc.bnl.gov/nsr/nsrlink.jsp?2017Lo03,B}{2017Lo03}&\parbox[t]{6in}{\addtolength{\parindent}{-0.25cm}A.E.Lovell, F.M.Nunes, I.J.Thompson - Phys.Rev. C 95, 034605 (2017).}\\
&\parbox[t]{6in}{\addtolength{\parindent}{-0.25cm} \textit{Three-body model for the two-neutron emission of \ensuremath{^{\textnormal{16}}}Be.}}\\
\href{https://www.nndc.bnl.gov/nsr/nsrlink.jsp?2018Ca09,B}{2018Ca09}&\parbox[t]{6in}{\addtolength{\parindent}{-0.25cm}J.Casal - Phys.Rev. C 97, 034613 (2018).}\\
&\parbox[t]{6in}{\addtolength{\parindent}{-0.25cm} \textit{Two-nucleon emitters within a pseudostate method: The case of \ensuremath{^{\textnormal{6}}}Be and \ensuremath{^{\textnormal{16}}}Be.}}\\
\href{https://www.nndc.bnl.gov/nsr/nsrlink.jsp?2018Fo07,B}{2018Fo07}&\parbox[t]{6in}{\addtolength{\parindent}{-0.25cm}H.T.Fortune - Eur.Phys.J. A 54, 51 (2018).}\\
&\parbox[t]{6in}{\addtolength{\parindent}{-0.25cm} \textit{Structure of exotic light nuclei: Z = 2, 3, 4.}}\\
\href{https://www.nndc.bnl.gov/nsr/nsrlink.jsp?2019Ca03,B}{2019Ca03}&\parbox[t]{6in}{\addtolength{\parindent}{-0.25cm}J.Casal, J.Gomez-Camacho - Phys.Rev. C 99, 014604 (2019).}\\
&\parbox[t]{6in}{\addtolength{\parindent}{-0.25cm} \textit{Identifying structures in the continuum: Application to \ensuremath{^{\textnormal{16}}}Be.}}\\
\href{https://www.nndc.bnl.gov/nsr/nsrlink.jsp?2019Fo09,B}{2019Fo09}&\parbox[t]{6in}{\addtolength{\parindent}{-0.25cm}H.T.Fortune - Phys.Rev. C 99, 044318 (2019).}\\
&\parbox[t]{6in}{\addtolength{\parindent}{-0.25cm} \textit{2n decays of \ensuremath{^{\textnormal{16}}}Be.}}\\
\href{https://www.nndc.bnl.gov/nsr/nsrlink.jsp?2020It02,B}{2020It02}&\parbox[t]{6in}{\addtolength{\parindent}{-0.25cm}N.Itagaki, T.Fukui, J.Tanaka, Y.Kikuchi - Phys.Rev. C 102, 024332 (2020).}\\
&\parbox[t]{6in}{\addtolength{\parindent}{-0.25cm} \textit{\ensuremath{^{\textnormal{8}}}He and \ensuremath{^{\textnormal{9}}}Li cluster structures in light nuclei.}}\\
\href{https://www.nndc.bnl.gov/nsr/nsrlink.jsp?2021Wa16,B}{2021Wa16}&\parbox[t]{6in}{\addtolength{\parindent}{-0.25cm}M.Wang, W.J.Huang, F.G.Kondev, G.Audi, S.Naimi - Chin.Phys.C 45, 030003 (2021).}\\
&\parbox[t]{6in}{\addtolength{\parindent}{-0.25cm} \textit{The AME 2020 atomic mass evaluation (II). Tables, graphs and references.}}\\
\href{https://www.nndc.bnl.gov/nsr/nsrlink.jsp?2022Gu11,B}{2022Gu11}&\parbox[t]{6in}{\addtolength{\parindent}{-0.25cm}J.Guo, D.H.Chen, X.-R.Zhou, Q.B.Chen, H.-J.Schulze - Chin.Phys.C 46, 064106 (2022).}\\
&\parbox[t]{6in}{\addtolength{\parindent}{-0.25cm} \textit{Effects of a kaonic meson on the ground-state properties of nuclei.}}\\
\href{https://www.nndc.bnl.gov/nsr/nsrlink.jsp?2022Yu02,B}{2022Yu02}&\parbox[t]{6in}{\addtolength{\parindent}{-0.25cm}Q.Yuan, S.Q.Fan, B.S.Hu, J.G.Li et al. - Phys.Rev. C 105, L061303 (2022).}\\
&\parbox[t]{6in}{\addtolength{\parindent}{-0.25cm} \textit{Deformed in-medium similarity renormalization group.}}\\
\href{https://www.nndc.bnl.gov/nsr/nsrlink.jsp?2023Mu11,B}{2023Mu11}&\parbox[t]{6in}{\addtolength{\parindent}{-0.25cm}B.Mukeru, M.B.Mahatikele, G.J.Rampho - Phys.Rev. C 107, 064313 (2023).}\\
&\parbox[t]{6in}{\addtolength{\parindent}{-0.25cm} \textit{Nuclear interactions in weakly bound neutron-rich nuclei.}}\\
\href{https://www.nndc.bnl.gov/nsr/nsrlink.jsp?2024Mo02,B}{2024Mo02}&\parbox[t]{6in}{\addtolength{\parindent}{-0.25cm}B.Monteagudo, F.M.Marques, J.Gibelin, N.A.Orr et al. - Phys.Rev.Lett. 132, 082501 (2024).}\\
&\parbox[t]{6in}{\addtolength{\parindent}{-0.25cm} \textit{Mass, Spectroscopy, and Two-Neutron Decay of \ensuremath{^{\textnormal{16}}}Be.}}\\
\end{longtable}
\end{document}